\documentclass[aps,pre,reprint,floatfix]{revtex4-2}

\usepackage{graphicx}
\usepackage{amsmath,amssymb}
\usepackage{booktabs}
\usepackage{xcolor}
\usepackage{placeins}
\usepackage{float}

\newcommand{\thY}{\theta_Y}
\newcommand{\dd}{\mathrm{d}}

\begin{document}

\title{Apparent contact angle equation for a dielectric droplet on a conductive hydrophilic surface in an electric field}

\author{Ren-Qiang Huang}
\affiliation{Fujian Provincial Key Lab for Soft Functional Materials Research,
Research Institute for Biomimetics and Soft Matter, Department of Physics,
College of Physical Science and Technology, Xiamen University,
Xiamen 361005, People's Republic of China}

\author{Gao-Xiao Jiang}
\affiliation{Fujian Provincial Key Lab for Soft Functional Materials Research,
Research Institute for Biomimetics and Soft Matter, Department of Physics,
College of Physical Science and Technology, Xiamen University,
Xiamen 361005, People's Republic of China}

\author{Zhong-Yi Li}
\affiliation{Fujian Provincial Key Lab for Soft Functional Materials Research,
Research Institute for Biomimetics and Soft Matter, Department of Physics,
College of Physical Science and Technology, Xiamen University,
Xiamen 361005, People's Republic of China}

\author{Rui Ma}
\affiliation{Fujian Provincial Key Lab for Soft Functional Materials Research,
Research Institute for Biomimetics and Soft Matter, Department of Physics,
College of Physical Science and Technology, Xiamen University,
Xiamen 361005, People's Republic of China}

\author{Chen-Xu Wu}
\email{cxwu@xmu.edu.cn}
\affiliation{Fujian Provincial Key Lab for Soft Functional Materials Research,
Research Institute for Biomimetics and Soft Matter, Department of Physics,
College of Physical Science and Technology, Xiamen University,
Xiamen 361005, People's Republic of China}

\date{\today}

\begin{abstract}
A dielectric droplet resting on a conductive substrate elongates when a
voltage is applied across the surrounding gap without touching
the upper electrode. We construct a closed form
theory of the equilibrium shapes by approximating the interface as a
truncated spheroid with a single shape parameter and obtain the dimensionless
electrostatic prefactor that corrects the free-space depolarization, based on the mirror symmetry across the substrate. Minimizing the resulting free energy along voltage
ramps reproduces the boundary-element-method (BEM) equilibria, both the droplet height
and the contact-line radius, with and without gravity, at negligible
computational cost. An explicit
apparent contact angle equation is found that collapses all the numerical calculation data. The configuration shows a behavior of fixed contact angle and an increasing apparent angle with the applied voltage, which is completely different from that of the Young--Lippmann equation.
\end{abstract}

\maketitle

\section{Introduction}

The interplay of electric fields and liquid interfaces has been
studied for more than a century, from droplet deformation and
disintegration in uniform fields~\cite{RN13,taylor1964,taylor1966,RN29,RN7}
to electrohydrodynamic patterning and jet
printing~\cite{RN34,RN32,RN31,RN6}; on substrates, electrically actuated
droplets underpin digital microfluidics~\cite{RN24,RN11}. For a
conductive droplet on a dielectric-coated electrode, the
Young--Lippmann equation~\cite{RN3,RN23} lowers the effective
solid--liquid tension, so the measured angle decreases with voltage, yet
the microscopic angle at the contact line retains the Young
value~\cite{young1805,RN4}, with only an apparent angle
responding~\cite{buehrle2003,mugele2007}.

The dual configuration---a dielectric droplet on a conductive substrate
in a parallel-plate gap it does not bridge---is qualitatively different:
the field penetrates and polarizes the liquid, the electrostatic energy
is stored in the droplet volume, and the primary response is bulk
elongation rather than a change in contact-angle~\cite{RN40,corson2014,vancauwenberghe2013}. Realized in non-contact
electrostatic manipulation~\cite{RN38,RN455}, this problem has been
addressed by experiments, asymptotics, and direct
simulation~\cite{corson2014,RN5,RN10,RN2}, but a closed-form theory of
the equilibrium shape is still lacking. This theory is desirable
because the configuration exhibits voltage-driven wetting transitions and
critical voltages~\cite{xiao2021,RN2,RN1} and a generic loss of
equilibrium in strong fields~\cite{RN33,RN18,RN26}, whose prediction
requires the coupled shape-field problem at every voltage.

\begin{figure}[htp]
  \centering
  \includegraphics[width=\linewidth,keepaspectratio]{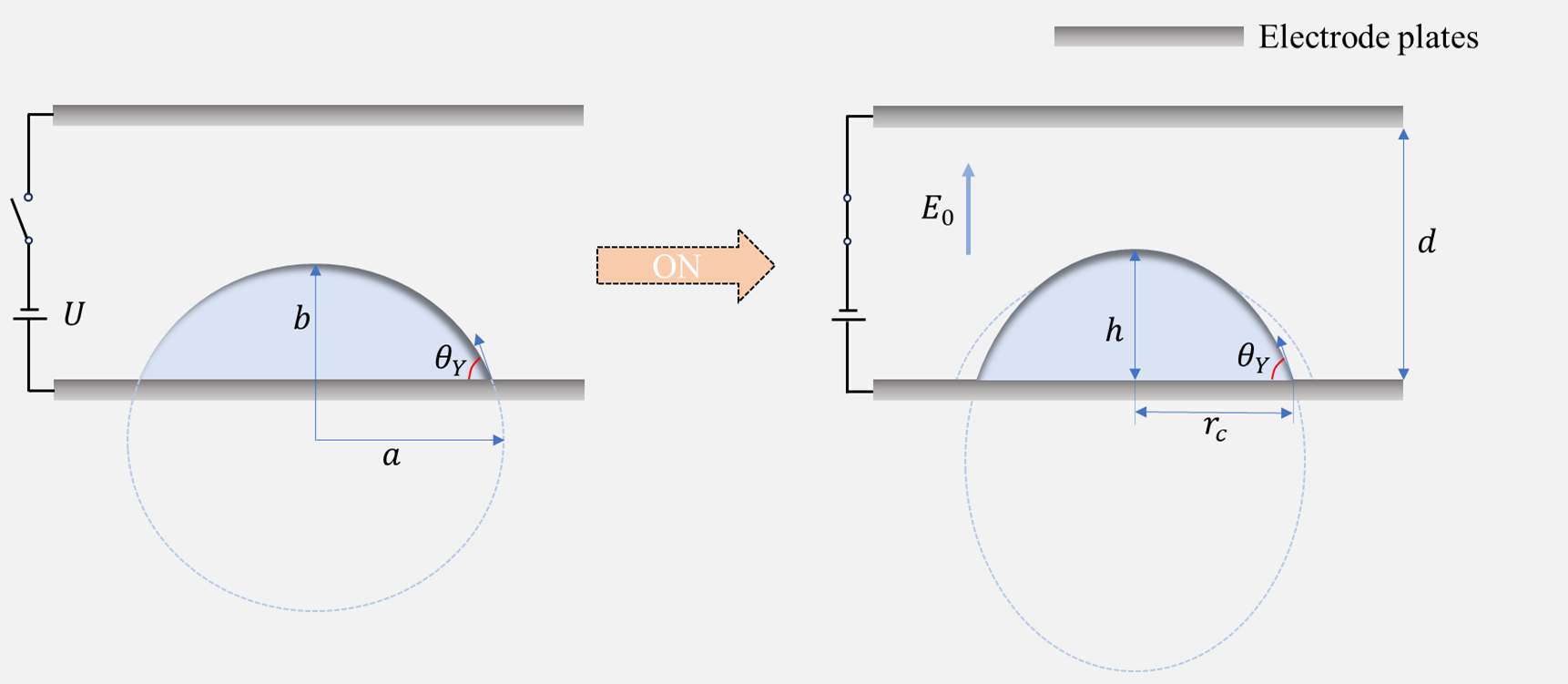}
  \caption{Schematic of the system. A dielectric droplet sits on the
  grounded lower plate of a parallel-plate capacitor of gap $d$; $a$
  and $b$ are the semi-axes of the truncated spheroid, $h$ the height,
  and $r_c$ the contact-line radius. Left, zero voltage; right, voltage
  $U$ applied. The contact angle stays at the Young value $\thY$ while
  the drop elongates in the field $E_0=U/d$; the droplet and its mirror
  image form the lens construction of Sec.~\ref{sec:prefactor}.
  \label{fig:schematic}}
\end{figure}

Here, we construct such a theory, with its proof relying on three observations. First,
the true contact angle is pinned at the Young angle $\thY$ by the Young
equation at the contact line~\cite{young1805,buehrle2003}, a fact we
verify with a variational model in which the contact angle is a free
variable (Sect.~\ref{sec:model}); the equilibrium shapes are thus
confined to a low-dimensional family, which truncated spheroids of
aspect ratio $k$ already capture quantitatively.
Second, the dimensionless electrostatic prefactor $f(\thY,k)$ of this
family has a closed form, obtained by mapping the droplet and its mirror
image onto a lens-shaped spheroid and correcting the free-space
depolarization factor; the result matches BEM
calculations within $2.7\%$ over the
investigated range $\pi/18\le\theta_Y\le2\pi/3$, $0.4\le k\le1.6$. The free energy is then explicit, with each
equilibrium found by one-dimensional minimization without solving any
field equation. Third, the theory predicts that an experimental measurement would yield
an apparent contact angle that varies with voltage and measurement scale,
even though the microscopic angle does not.

The paper is organized as follows. Section~\ref{sec:model} defines the
model, the free-energy construction, and the two numerical
implementations used in this work. Section~\ref{sec:prefactor} derives
the closed-form electrostatic prefactor. Section~\ref{sec:shapes}
compares the theoretical equilibria with the BEM solutions and maps the
equilibrium shape on the $(\thY,U)$ plane.
Section~\ref{sec:apparent} establishes the apparent-angle scaling law,
and Sec.~\ref{sec:conclusion} concludes.

\section{Model and methods}
\label{sec:model}

We consider an axisymmetric dielectric droplet of volume $V_0$,
surface tension $\sigma$, density contrast $\Delta\rho$ and relative
permittivity $\varepsilon_r$, sessile on the lower plate of a
parallel-plate capacitor of gap $d$, across which a voltage $U$ is
applied. Here $\varepsilon_r$ is a material parameter; the results
below use $\varepsilon_r=80$ (water), and $\varepsilon_r=40$ is used in
Fig.~\ref{fig:apparent} to test the generality of the scaling law [Fig.~\ref{fig:schematic}]. The far field is uniform,
$E_0=U/d$, and the Bond number $\mathrm{Bo}=\Delta\rho gV_0^{2/3}/\sigma$ reflects the ratio of the contributions made by gravity and surface tension; we report results with and
without gravity throughout. Throughout the paper, lengths are measured in
units of $V_0^{1/3}$, and the applied voltage is measured in units of the characteristic
voltage
$U^*=d\sqrt{2\sigma/[\varepsilon_0(\varepsilon_r-1)V_0^{1/3}]}$.

At the sharp-interface level, the contact line is a boundary condition
rather than a variational degree of freedom. The Young equation fixes
the microscopic contact angle at $\thY$~\cite{young1805}, and the
electrostatic contribution to the contact-line force vanishes because
the wedge singularity of the field is
integrable~\cite{buehrle2003,mugele2007}. We test this assumption
independently with a variational (spectral) model of the full profile
in which the contact angle is a free optimization variable: minimizing
the total energy over $0\le U/U^*\le4.1$, returns $\theta_c=\thY$
within $0.6^\circ$, while the height increases by up to $22\%$. The
free-boundary BEM solver described in the following therefore imposes
$\theta_c=\thY$ as its contact-line boundary condition and serves as
the benchmark for the shape. The equilibrium problem therefore
reduces to shape variations at fixed $\thY$. Guided by this fact, we approximate the
interface by a truncated spheroid, i.e., the cap of an ellipsoid of
equatorial and polar semi-axes $a$ and $b$, cut by the substrate such
that the contact angle is equal to $\thY$. 

In the DC voltage ensemble, the total free energy reads
\begin{equation}
\begin{split}
F={}&-\frac{1}{2}\varepsilon_0 E_0^2 V_0\, f(\thY,k)
    + \sigma V_0^{2/3} A(\thY,k)\\
    &+ \Delta\rho g V_0^{4/3} B(\thY,k),
\end{split}
\label{eq:F}
\end{equation}
where $f$, $A$, $B$ are dimensionless prefactors of the electrostatic,
surface, and gravitational contributions. The geometric prefactors $A$
and $B$ are analytic functions of the ellipsoid geometry and are given
in Appendix~\ref{app:geom}. The electrostatic prefactor $f$ is the
nontrivial input and is derived in Sec.~\ref{sec:prefactor}.

In this work, two numerical implementations are used. The first is the
theory itself. At each voltage, the volume constraint
[Eq.~(\ref{eq:vol})] fixes the scale $b$ for a given $k$, so that the
free energy Eq.~(\ref{eq:F}) becomes a function of $k$ alone; the equilibrium is located by the bounded one-dimensional minimization of the free energy with respect to $k$, followed by a continuation in $U$ from the zero-voltage state, with the
converged $k$ at one voltage serving as the initial guess in the next.

The second implementation is an independent sharp-interface
free-boundary solver that serves as the benchmark. The electric problem
is solved by an axisymmetric boundary-element method. The potential is
represented as $\varphi=\varphi_0+\oint_{\Gamma}
G(\mathbf{x},\mathbf{x}')\sigma_{\rm p}(\mathbf{x}')\,\dd S'$ with the applied
uniform potential $\varphi_0=E_0 z$, and the interface conditions of
continuous $\varphi$ and $\varepsilon\,\partial\varphi/\partial n$
reduce to a second-kind Fredholm equation for the polarization charge
density $\sigma_{\rm p}$,
\begin{equation}
\sigma_{\rm p}(\mathbf{x}) + 2\lambda\varepsilon_0
\Bigl[\oint_{\Gamma}\frac{\partial G}{\partial n}(\mathbf{x},\mathbf{x}')
\sigma_{\rm p}(\mathbf{x}')\,\dd S'
+ \frac{\partial\varphi_0}{\partial n}(\mathbf{x})\Bigr] = 0,
\label{eq:bem}
\end{equation}
where $\lambda=(\varepsilon_r-1)/(\varepsilon_r+1)$, $G$ is the
Green's half-space function of the grounded substrate, and the
azimuthally integrated kernels are expressed through complete elliptic
integrals. The interface is discretized into arc-length
segments with a piecewise-linear charge density, collocated at interior
Gauss points to avoid the spurious singularity of the segment normals.
From the solution, the Maxwell stress along the outward normal is
\begin{equation}
f_n=\frac{1}{2}\varepsilon_0(\varepsilon_r-1)
\left(\varepsilon_r E_n^2+E_t^2\right),
\label{eq:maxwell}
\end{equation}
with $E_n$ and $E_t$ the normal and tangential field components on the
liquid side. The electrostatic free energy, which defines the prefactor
$f$ of Eq.~(\ref{eq:F}), is evaluated from the polarization charge as
\begin{equation}
W_e=-\frac12\int_V \mathbf P\cdot\mathbf E_0\,\dd V
=-\frac12\varepsilon_0 E_0^2 V_0 f,
\label{eq:we}
\end{equation}
where $\int_V\mathbf P\cdot\mathbf E_0\,\dd V
=\oint_\Gamma\sigma_{\rm p}\varphi_0\,\dd S$ with the applied potential
$\varphi_0=E_0 z$, so that the BEM solution determines
$f=\oint_\Gamma\sigma_{\rm p}\varphi_0\,\dd S/(\varepsilon_0 E_0^2 V_0)$.
The shape obeys the Young--Laplace equation in a strong form,
integrated as an initial-value problem in arc length $s$ from the apex,
\begin{equation}
\begin{split}
\frac{\dd r}{\dd s}&=\cos\psi,\qquad
\frac{\dd z}{\dd s}=\sin\psi,\\
\frac{\dd\psi}{\dd s}
&=-\frac{\Delta p(z)+f_n(s)}{\sigma}-\frac{\sin\psi}{r},
\end{split}
\label{eq:yl}
\end{equation}
where $\psi$ is the tangent angle and
$\Delta p(z)=\Delta p_0+\Delta\rho g(h-z)$ is the local pressure jump.
The three shooting parameters, namely the apex height $h$, the pressure
jump $\Delta p_0$, and the arc length of the contact line $L$, are fixed by the
conditions $z(L)=0$, $\psi(L)=-\thY$, and $V=V_0$ at the contact line.
The field and shape are iterated with
under-relaxation until self-consistent, and the branch is built by
continuation in $U$. The solver is validated against the classical
depolarization result for spheres and spheroids ($0.08$--$0.25\%$) and
against independent finite-element (COMSOL) energies (to within
$0.3\%$), as detailed in Appendix~\ref{app:errors}.

\section{Closed-form electrostatic prefactor}
\label{sec:prefactor}

For a complete spheroid in a uniform far field, the polarization energy
is classical~\cite{landau8},
\begin{equation}
f=\frac{\varepsilon_r-1}{1+N(\varepsilon_r-1)},
\label{eq:f}
\end{equation}
with $N=n_z(k)$ the depolarization factor,
\begin{equation}
n_z(k)=
\begin{cases}
\dfrac{k^2}{e^3}\bigl(\operatorname{arctanh}e-e\bigr),
& k\le 1,\ e=\sqrt{1-k^2},\\[8pt]
\dfrac{k^2}{e^3}\bigl(e-\arctan e\bigr),
& k\ge 1,\ e=\sqrt{k^2-1},
\end{cases}
\label{eq:nz}
\end{equation}
which equals $1/3$ for a sphere. For a truncated spheroid on a
conductive substrate, such a closed form does not exist.

\begin{figure}[htp]
  \centering
  \includegraphics[width=\linewidth,keepaspectratio]{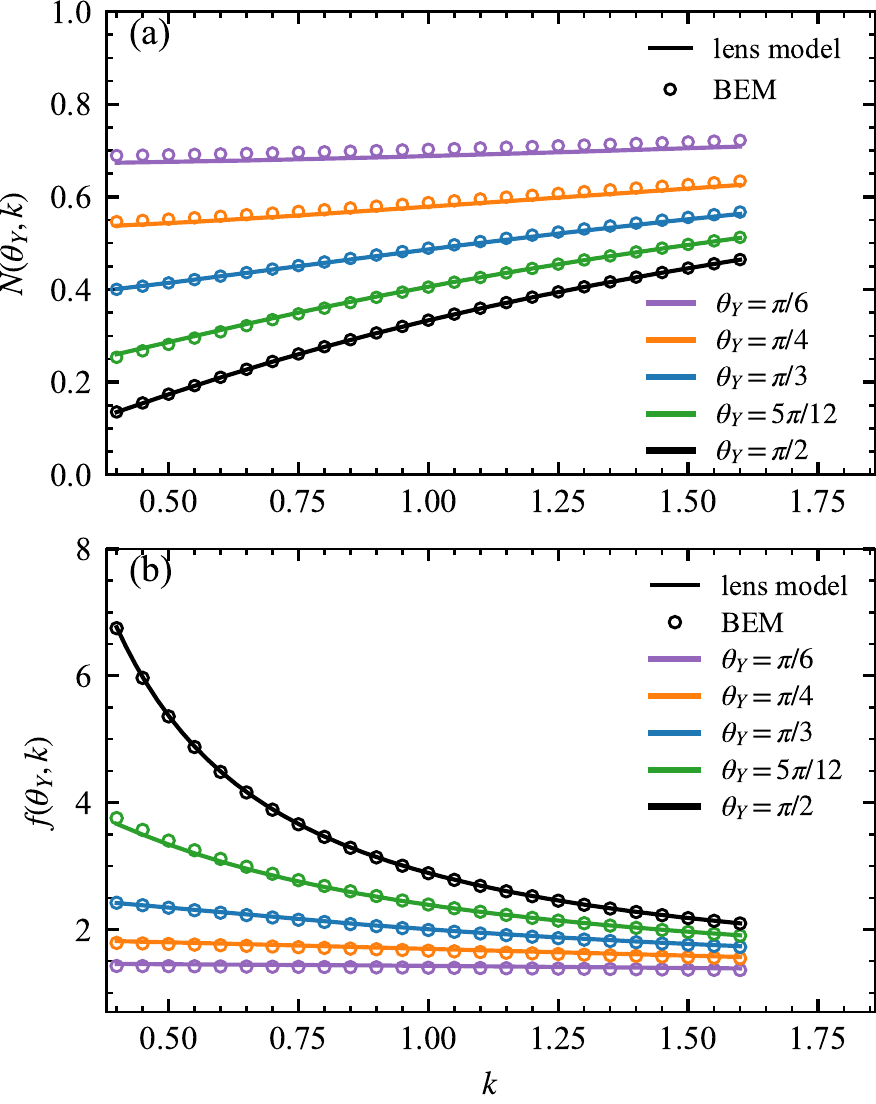}
  \caption{(a) Effective depolarization factor $N(\thY,k)$ given by the
  lens model, Eqs.~(\ref{eq:klens}) and (\ref{eq:S}) (solid lines),
  compared with BEM calculations (open circles) for five
  Young angles. (b) Electrostatic prefactor $f(\thY,k)$ given by the
  lens model (solid lines), compared with the same BEM
  data.
  \label{fig:prefactor}}
\end{figure}

The grounded substrate acts as a mirror, so the droplet and its image
form a double-capped lens of aspect ratio
\begin{equation}
k_{\rm lens}=k\,\frac{\sqrt{1-\zeta^2}}{1+\zeta},\qquad
\zeta=-\frac{\operatorname{sgn}(\cos\thY)}{\sqrt{1+k^2\tan^2\thY}}.
\label{eq:klens}
\end{equation}

We take the effective depolarization factor of the truncated spheroid as
\begin{equation}
N(\thY,k)=n_z(k_{\rm lens}),
\label{eq:S}
\end{equation}
which at $\thY=\pi/2$ recovers $N=n_z(k)$ and the half-spheroid value $1/3$ exactly. 

Figure~\ref{fig:prefactor} compares Eqs.~(\ref{eq:klens}) and (\ref{eq:S})
with BEM calculations of $f$ on a $25\times 25$ grid spanning $\thY=\pi/6$--$\pi/2$ and $k=0.4$--$1.6$, showing that the lens model reproduces well both the depolarization factor and the prefactor over the whole range. 

The BEM values agree
with the independent finite-element (COMSOL) energies to within $0.3\%$ (see the Appendix), indicating that the residual error is attributable to the model rather than the numerics.

With $f$ given by Eqs.~(\ref{eq:f}), (\ref{eq:klens}) and (\ref{eq:S})
and $A$, $B$ analytic (Appendix~\ref{app:geom}), the entire free energy
Eq.~(\ref{eq:F}) becomes explicit and no field equation is needed to be solved at any stage of the equilibrium calculation. 


\section{Equilibrium shapes}
\label{sec:shapes}

We validated the theory against BEM free-boundary solutions on 32 equilibrium branches spanning $\thY=\pi/6$--$\pi/2$, with and without gravity,
each followed by a continuation in $U$. 

\begin{figure}[htp]
  \centering
  \includegraphics[width=\linewidth,keepaspectratio]{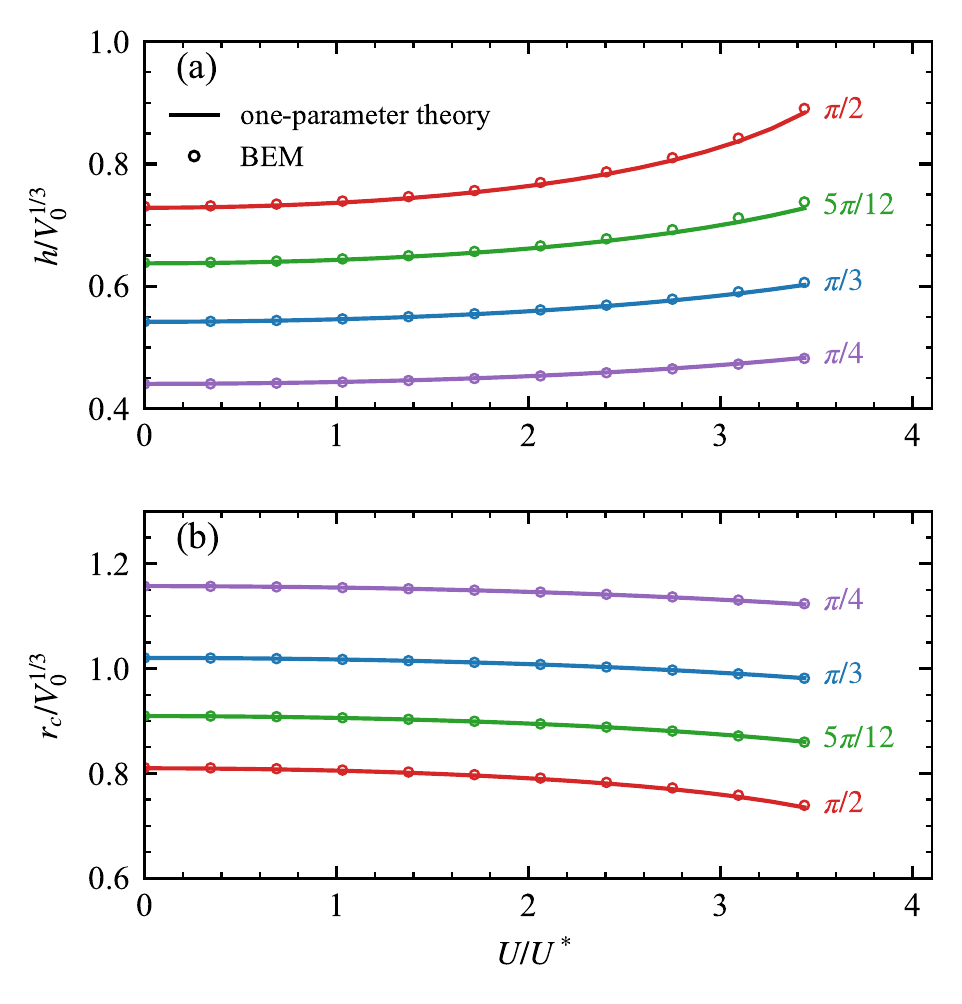}
  \caption{(a) Equilibrium height $h/V_0^{1/3}$ and (b) contact-line radius
  $r_c/V_0^{1/3}$ as functions of the dimensionless voltage $U/U^*$ at
  $\mathrm{Bo}=0.82$ and $d/V_0^{1/3}\simeq2.03$ for
  $\thY=\pi/4,\pi/3,5\pi/12,\pi/2$, comparing the theory (solid lines) with the BEM free-boundary
  solutions (open circles).
  \label{fig:chains}}
\end{figure}

Figure~\ref{fig:chains} shows four
representative branches at $\mathrm{Bo}=0.82$, indicating that the theory
closely follows the benchmark for height and radius of the contact line.

The precision is controlled by the prefactor rather than the shape
family: recomputing with the uncorrected free-space factor $n_z(k)$
degrades the height error to $8\%$ at large angles, and the branches
terminate $0.3$--$0.7\,U^*$ early. The cost is negligible---each
equilibrium is a one-dimensional minimization of an analytic function,
whereas the benchmark needs tens of coupled free-boundary solves per
voltage.

\begin{figure}[htp]
  \centering
  \includegraphics[width=\linewidth,keepaspectratio]{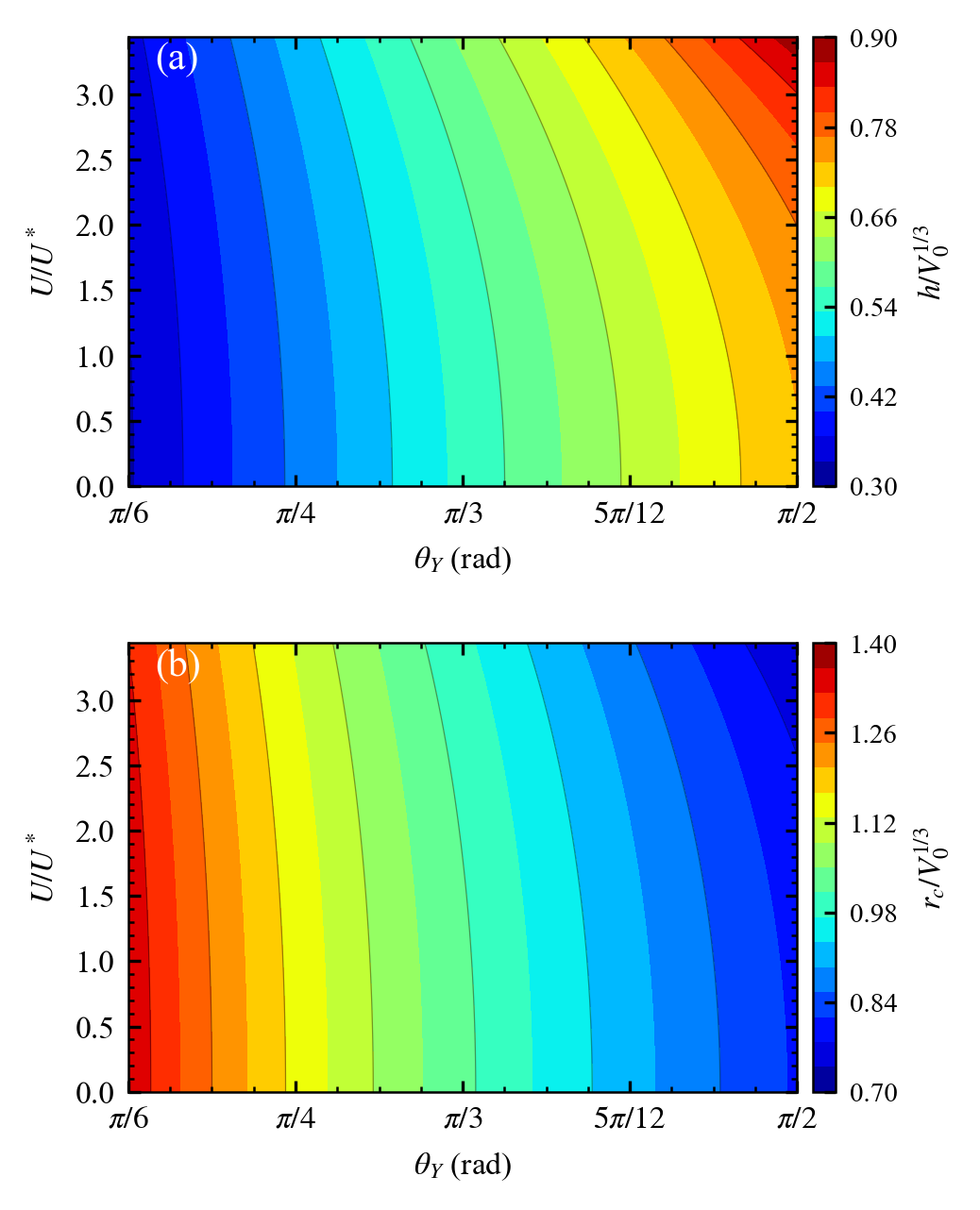}
  \caption{Equilibrium shape over the parameter plane at
  $\mathrm{Bo}=0.82$ and $d/V_0^{1/3}\simeq2.03$: (a) droplet height
  $h/V_0^{1/3}$ and (b) contact-line radius $r_c/V_0^{1/3}$ as functions of
  the Young angle $\thY$ and the dimensionless voltage $U/U^*$.
  \label{fig:phase}}
\end{figure}
Figure~\ref{fig:phase} assembles the validated theory into a map of the
equilibrium shape on the $(\thY,U)$ plane with gravity. The response
strongly depends on the Young angle. At $\thY=\pi/3$, increasing
$U/U^*$ to $3.4$ raises the height by $12\%$ and reduces the radius of
the contact line by $4\%$, while at $\thY=\pi/2$ the same ramp raises
the height by $22\%$ and shrinks the contact line by $9\%$. In this whole map, the theory deviates
from the benchmark by no more than the few percent documented in
Fig.~\ref{fig:errors}. The largest deviations occur near the
upper-right corner, where the equilibrium branch approaches its
terminal turning point. No adjustable parameter enters either map.

\section{Apparent contact angle equation}
\label{sec:apparent}

The pinned contact angle does not imply a null measurement: a
goniometer fits the visible profile, which does change. In
electrowetting, the measured angle is the spherical cap sharing the
drop's volume and contact radius, extrapolated to the substrate; the
two definitions coincide because the field distorts only a small region
near the contact line~\cite{buehrle2003,mugele2007}. Here, the distortion
spans the whole profile, so no unique macroscopic angle exists.

We therefore define the local tangent angle $\theta_{\rm app}(\ell)$ at
the arc length $\ell$ from the contact line and the field-induced change
$\delta\theta_{\rm app}(\ell;U)=\theta_{\rm app}(\ell;U)
-\theta_{\rm app}(\ell;0)$ relative to the zero-voltage profile. We also
introduce the cosine change
$\delta\cos\theta_{\rm app}(\ell;U)\equiv\cos\theta_{\rm app}(\ell;U)-\cos\theta_{\rm app}(\ell;0)$.

The change in $\theta_{\rm app}(\ell)$ is positive, linear in $\ell$ in the observable range, and
increases quadratically with $U$. Integrating the Young--Laplace equation
[Eq.~(\ref{eq:yl})] from the contact line gives
\begin{equation}
\theta_{\rm app}(\ell)
=\thY-\frac{1}{\sigma}\int_0^\ell
\bigl[\Delta p(z)+f_n(\ell')\bigr]\,\dd\ell'
+\int_0^\ell\frac{\sin\theta_{\rm app}(\ell')}{r(\ell')}\,\dd\ell',
\label{eq:appint}
\end{equation}
At zero voltage, the profile is a spherical cap, on which
$\sin\theta_{\rm app}/r=\sin\thY/R_0$ is constant. Near the contact line
($\ell\ll R_0$) the azimuthal-curvature term therefore changes only at
$O(\ell/R_0)$ relative to the leading electrostatic term and is
removed in what follows. Subtracting the zero-voltage profile and
linearizing the cosine at the zero-field tangent angle, in leading order
$\delta\cos\theta_{\rm app}(\ell;U)=-\sin\thY\,\delta\theta_{\rm app}(\ell)$,
and all data collapse onto
\begin{equation}
\delta\cos\theta_{\rm app}(\ell;U)
=-c(\thY)\,\sin\thY\,\Gamma\,\frac{\ell}{R_0},
\label{eq:apparent}
\end{equation}
where $\Gamma=\varepsilon_0(\varepsilon_r-1)U^2R_0/(2\sigma d^2)$ is the
electro-capillary number and $R_0=\left\{3/[\pi(2-3\cos\thY+\cos^3\thY)]\right\}^{1/3}V_0^{1/3}$ is the radius of curvature of the zero-field spherical cap. The dependence of $c(\thY)$ on the
Young angle can be fixed as follows. In the contact line limit, the
balance of electrostatic driving against capillary stiffness gives
$\delta k\propto U^2 f_k/\mathcal F_0''$, where
$f_k\equiv\partial f/\partial k$ and $\mathcal F_0''$ is the second
derivative of the zero-field free energy with respect to $k$; the
resulting change in the meridional curvature of the contact line gives
$\delta\theta_{\rm app}\propto U^2 f_k/\mathcal F_0''$. Combined with
Eq.~(\ref{eq:apparent}), this yields
\begin{equation}
c(\thY)\propto\frac{f_{\rm sph}^2}{\varepsilon_r-1},
\end{equation}
where
$f_{\rm sph}=f(\pi/2,1)=(\varepsilon_r-1)/(1+(\varepsilon_r-1)/3)$ is
the half-spheroid prefactor. A least-squares fit of the BEM data over
$\ell/R_0\in[0.015,0.12]$ fixes the remaining numerical coefficient to
$(1-\cos\thY)/2$, so that at acute and right angles,

\begin{widetext}
\begin{equation}
\delta\cos\theta_{\rm app}(\ell;U)
=-\frac{9(\varepsilon_r-1)}{2(\varepsilon_r+2)^2}(1-\cos\thY)
\sin\thY\,\Gamma\,\frac{\ell}{R_0}.
\label{eq:apparent contact angle}
\end{equation}
\end{widetext}

\begin{figure}[htp]
  \centering
  \includegraphics[width=\linewidth,keepaspectratio]{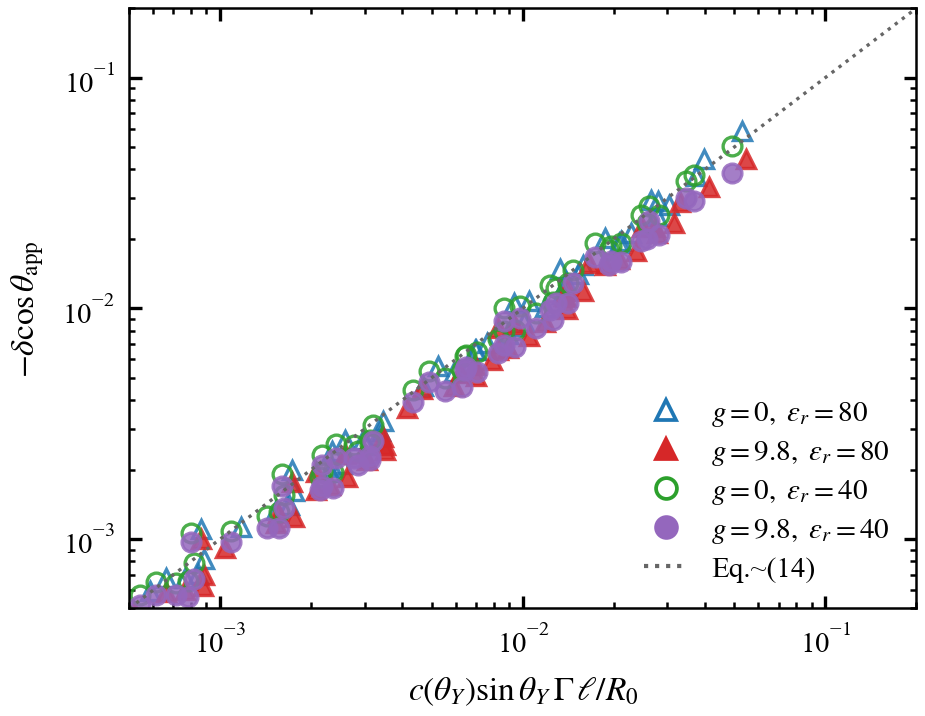}
  \caption{Collapse of different parameter combinations onto the scaling
  law Eq.~(\ref{eq:apparent contact angle}) (dotted line), for
  $\theta_Y=\pi/4,\pi/3,5\pi/12,\pi/2$ and $\ell/R_0\in[0.03,\,0.12]$, and
  $U/U^*\in[0.65,\,3.4]$.
  Markers: open triangles ($\mathrm{Bo}=0$, $\varepsilon_r=80$), filled
triangles ($\mathrm{Bo}=0.82$, $\varepsilon_r=80$), open circles
($\mathrm{Bo}=0$, $\varepsilon_r=40$), and filled circles
($\mathrm{Bo}=0.82$, $\varepsilon_r=40$). $R_0$ in $\Gamma$, if scaled by $V_0^{1/3}$, can be uniquely determined by $\theta_Y$.}
  \label{fig:apparent}
\end{figure}

\begin{figure}[H]
  \centering
  \includegraphics[width=\linewidth,keepaspectratio]{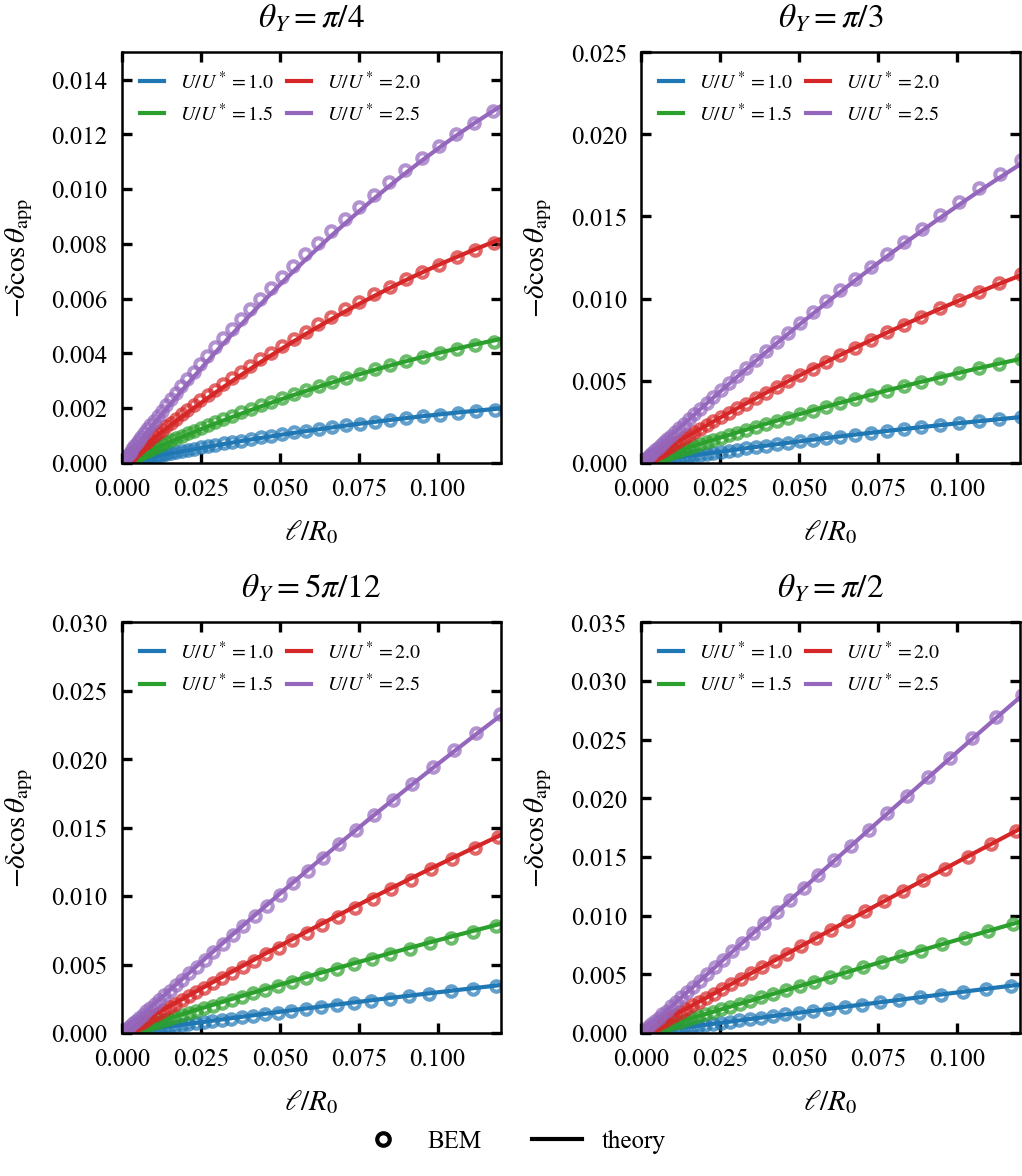}
  \caption{Cosine change of the local tangent angle
  $-\delta\cos\theta_{\rm app}(\ell)$ versus $\ell/R_0$ at $\mathrm{Bo}=0$ for four Young angles $\theta_Y=\pi/4,\pi/3,5\pi/12,\pi/2$ and four voltages
  $U/U^*=1.0,1.5,2.0,2.5$ each, the color encoding the voltage,
  comparing the theory (solid lines) with the BEM free-boundary
  benchmark (open circles).
  \label{fig:appcompare}}
\end{figure}

Figure~\ref{fig:apparent} shows that different combinations of
parameters collapse onto the scaling law
Eq.~(\ref{eq:apparent contact angle}), establishing it as a numerically
established scaling law. Rescaling is insensitive to gravity: the $\mathrm{Bo}=0.82$ data
(dashed) fall onto the same master curve as the $\mathrm{Bo}=0$ data
within the investigated range. Figure~\ref{fig:appcompare}
further compares the theory with the BEM free-boundary benchmark for
the cosine change $-\delta\cos\theta_{\rm app}(\ell)$ at four Young's contact
angles, confirming that Eq.~(\ref{eq:apparent contact angle}) describes
the apparent contact angle of a dielectric droplet on a conductive
hydrophilic surface in an electric field.

\section{Conclusion}
\label{sec:conclusion}

We have shown that the equilibrium of a sessile dielectric droplet in a
non-contact electric field admits an accurate closed-form description by approximating the interface as a truncated spheroid with a fixed contact angle and a single shape parameter. The dimensionless electrostatic prefactor corrects the free-space polarization well. Based on this, an explicit apparent contact angle equation is found that does not depend on gravity and collapses all the numerical calculation data. It also demonstrates a behavior different from that of the Young-Lippmann equation that increases with the applied voltage.

\begin{acknowledgments}
We acknowledge the financial support from the China National Natural Science Foundation under Grant No.~12574232 and the 111 project B16029.
\end{acknowledgments}

The data supporting the findings of this article are available from
the authors upon reasonable request.

\appendix

\section{Analytic geometric prefactors}
\label{app:geom}

For the truncated-spheroid family, the geometry is fixed by the aspect
ratio $k=a/b$, the truncation parameter $\zeta$ defined in
Eq.~(\ref{eq:klens}), and the volume constraint. With the vertical coordinate measured in units of $b$ from the spheroid
center, the cap spans $u\in[-\zeta,1]$, and the volume constraint $V=V_0$ is given by
\begin{equation}
V_0=\pi k^2 b^3 \mathcal{F}(\zeta),\qquad
\mathcal{F}(\zeta)=\frac{2}{3}+\zeta-\frac{\zeta^3}{3},
\label{eq:vol}
\end{equation}
so that $b=\left[V_0/(\pi k^2\mathcal{F})\right]^{1/3}$ and $a=kb$.

The liquid--gas area of the cap is
\begin{equation}
\begin{split}
A_{\rm LG}&=2\pi ab\int_{-\zeta}^{1}\sqrt{1-(1-k^2)u^2}\,\dd u\\
&=2\pi ab\,\bigl[\mathcal{I}(1)-\mathcal{I}(-\zeta)\bigr],
\end{split}
\label{eq:alg}
\end{equation}
with the elementary antiderivative
\begin{equation}
\mathcal{I}(u)=
\begin{cases}
\dfrac{1}{2}\left[u\sqrt{1-e^2u^2}
+\dfrac{\arcsin(eu)}{e}\right],
& k\le 1,\\[10pt]
\dfrac{1}{2}\left[u\sqrt{1+e^2u^2}
+\dfrac{\operatorname{arcsinh}(eu)}{e}\right],
& k\ge 1,
\end{cases}
\label{eq:anti}
\end{equation}
where $e=\sqrt{|1-k^2|}$ and $\mathcal{I}(u)=u$ for $k=1$. The solid--liquid area is
$A_{\rm SL}=\pi a^2(1-\zeta^2)$, and the surface prefactor of
Eq.~(\ref{eq:F}) is
\begin{equation}
A(\thY,k)=V_0^{-2/3}\left(A_{\rm LG}-\cos\thY\,A_{\rm SL}\right).
\label{eq:A}
\end{equation}

The centroid height above the substrate is
\begin{equation}
\begin{split}
&z_c=\frac{\pi a^2 b^2}{V_0}\bigl[\mathcal{P}(1)-\mathcal{P}(-\zeta)\bigr],\\
&\mathcal{P}(u)=\zeta\left(u-\frac{u^3}{3}\right)+\frac{u^2}{2}
-\frac{u^4}{4},
\end{split}
\label{eq:zc}
\end{equation}
and the gravitational prefactor is $B(\thY,k)=z_c/V_0^{1/3}$.

\section{Accuracy of the closed-form theory}
\label{app:errors}

\begin{figure}[htp]
  \centering
  \includegraphics[width=\linewidth,keepaspectratio]{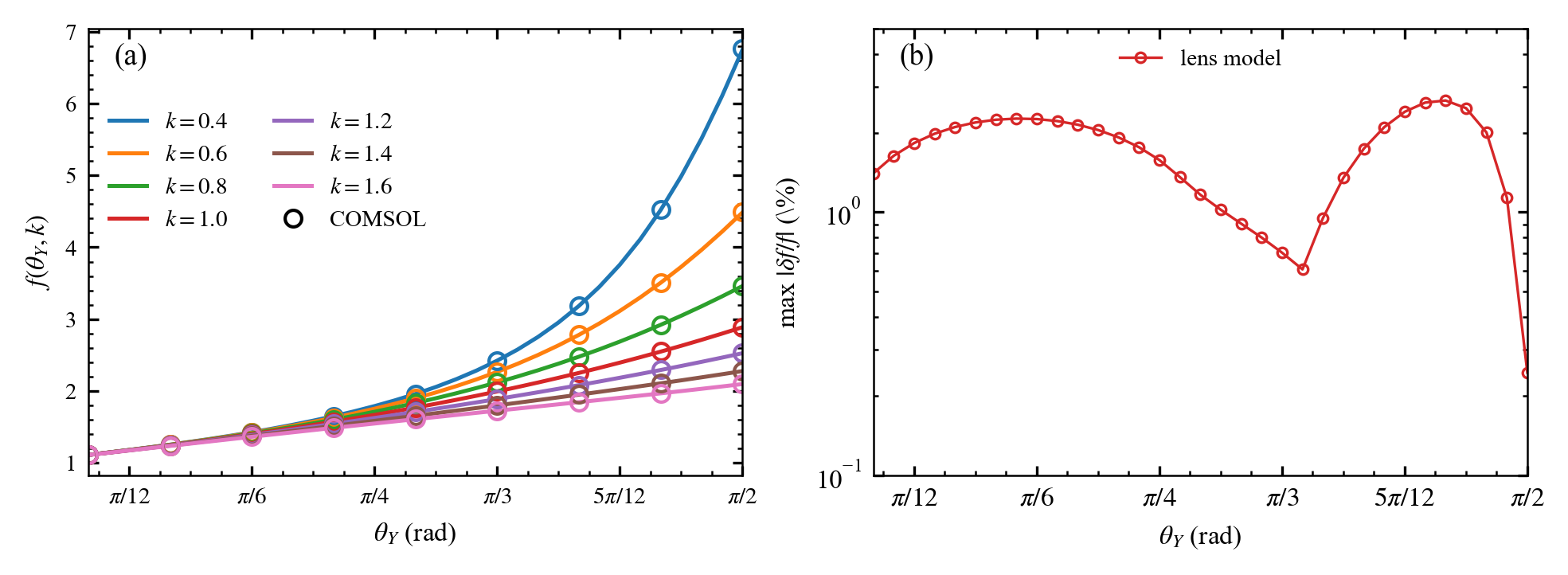}
  \caption{(a) Validation of the BEM electrostatic prefactor
  $f(\thY,k)$ against independent finite-element (COMSOL) calculations.
  Solid lines of different colors: BEM for $k=0.4,\dots,1.6$; open
  circles of the same color: COMSOL at the same $(\thY,k)$ points over
  $\thY=\pi/18$--$\pi/2$. The largest difference is $0.3\%$.
  (b) Maximum relative error of the electrostatic prefactor $f$ over the
  sampled $(\thY,k)$ plane for the lens model.}
  \label{fig:comsol}
\end{figure}

The BEM solver is validated against independent finite-element (COMSOL)
calculations. Figure~\ref{fig:comsol}(a) compares the electrostatic
prefactor $f$ obtained by the two methods on $\theta_Y=\pi/18$--$\pi/2$
and $k=0.4$--$1.6$; the largest relative difference is $0.3\%$.
Figure~\ref{fig:comsol}(b) reports the maximum relative error of the
prefactor $f$ on the $25\times25$ BEM grid of Sec.~\ref{sec:prefactor}:
the lens model stays at or below $2.7\%$ for every Young angle. The
remaining equilibrium-shape error is small and, at acute angles, is
dominated by the prefactor rather than the truncated-spheroid family.

\begin{figure}[htp]
  \centering
  \includegraphics[width=0.90\linewidth,keepaspectratio]{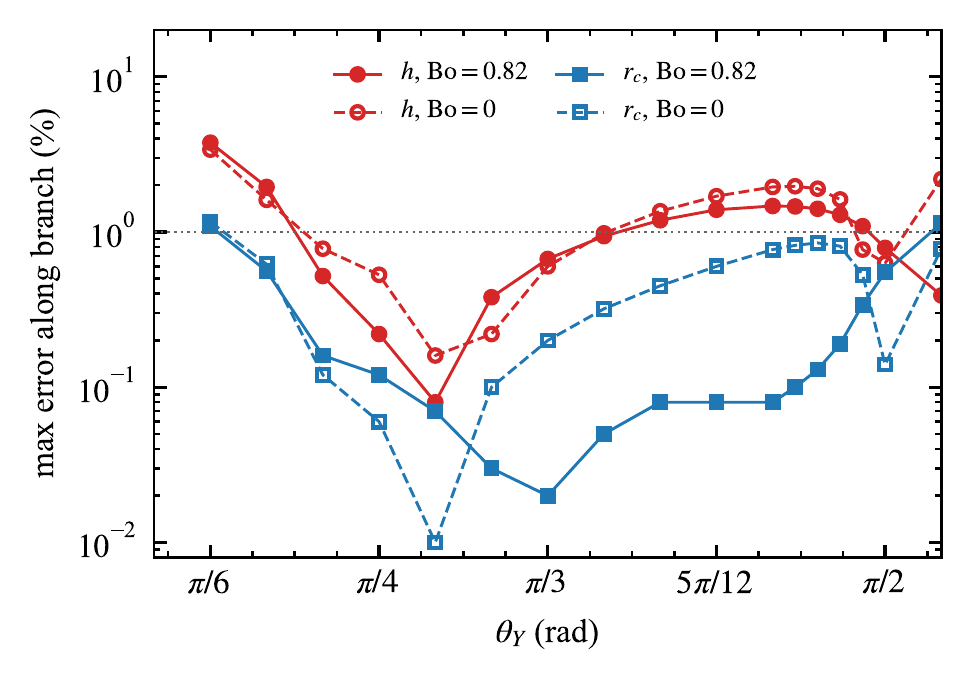}
  \caption{Maximum relative error along the equilibrium branch versus
  $\thY$ for the height $h$ (circles) and the contact-line radius $r_c$
  (squares), with (filled symbols) and without (open symbols) gravity;
  the dotted line marks $1\%$.}
  \label{fig:errors}
\end{figure}

Figure~\ref{fig:errors} reports the maximum relative error of height and
contact-line radius along each of the 32 branches of Sec.~\ref{sec:shapes},
with and without gravity. The height error stays below $2\%$ for
$\thY=7\pi/36$--$\pi/2$ and grows only near the highest angle termination
points, the radius of the contact line following the same trend with
smaller errors at acute angles.

\bibliographystyle{apsrev4-2}
\bibliography{myref}

@article{RN13,
   author = {O'Konski, C. T. and Thacher, H. C.},
   title = {The Distortion of Aerosol Droplets by an Electric Field},
   journal = {The Journal of Physical Chemistry},
   volume = {57},
   number = {9},
   pages = {955-958},
   DOI = {10.1021/j150510a024},
   year = {1953}
}

@article{taylor1964,
   author = {Taylor, G. I.},
   title = {Disintegration of water drops in an electric field},
   journal = {Proceedings of the Royal Society of London. Series A},
   volume = {280},
   number = {1382},
   pages = {383-397},
   year = {1964}
}

@article{taylor1966,
   author = {Taylor, G. I. and McEwan, A. D. and de Jong, L. N. J.},
   title = {Studies in electrohydrodynamics. I. The circulation produced in a drop by an electric field},
   journal = {Proceedings of the Royal Society of London. Series A},
   volume = {291},
   number = {1425},
   pages = {159-166},
   DOI = {10.1098/rspa.1966.0086},
   year = {1966}
}

@article{RN29,
   author = {Cheng, K. J. and Chaddock, J. B.},
   title = {Deformation and stability of drops and bubbles in an electric field},
   journal = {Physics Letters A},
   volume = {106},
   number = {1-2},
   pages = {51-56},
   DOI = {10.1016/0375-9601(84)90491-2},
   year = {1984}
}

@article{RN7,
   author = {Adamiak, K. and Floryan, J. M.},
   title = {Dynamics of Water Droplet Distortion and Breakup in a Uniform Electric Field},
   journal = {IEEE Transactions on Industry Applications},
   volume = {47},
   number = {6},
   pages = {2374-2383},
   DOI = {10.1109/TIA.2011.2168797},
   year = {2011}
}

@article{RN6,
   author = {Vlahovska, Petia M.},
   title = {Electrohydrodynamics of Drops and Vesicles},
   journal = {Annual Review of Fluid Mechanics},
   volume = {51},
   pages = {305-330},
   DOI = {10.1146/annurev-fluid-122316-050120},
   year = {2019}
}

@article{RN3,
   author = {Lippmann, G.},
   title = {Relations entre les ph\'enom\`enes \'electriques et capillaires},
   journal = {Annales de Chimie et de Physique},
   volume = {5},
   pages = {494-549},
   year = {1875}
}

@article{young1805,
   author = {Young, T.},
   title = {An essay on the cohesion of fluids},
   journal = {Philosophical Transactions of the Royal Society of London},
   volume = {95},
   pages = {65-87},
   year = {1805}
}

@article{RN4,
   author = {de Gennes, P. G.},
   title = {Wetting: statics and dynamics},
   journal = {Reviews of Modern Physics},
   volume = {57},
   number = {3},
   pages = {827-863},
   DOI = {10.1103/RevModPhys.57.827},
   year = {1985}
}

@article{RN23,
   author = {Mugele, F. and Baret, J. C.},
   title = {Electrowetting: From basics to applications},
   journal = {Journal of Physics: Condensed Matter},
   volume = {17},
   number = {28},
   pages = {R705-R774},
   DOI = {10.1088/0953-8984/17/28/r01},
   year = {2005}
}

@article{buehrle2003,
   author = {Buehrle, J. and Herminghaus, S. and Mugele, F.},
   title = {Interface profiles near three-phase contact lines in electric fields},
   journal = {Physical Review Letters},
   volume = {91},
   number = {8},
   pages = {086101},
   DOI = {10.1103/PhysRevLett.91.086101},
   year = {2003}
}

@article{mugele2007,
   author = {Mugele, F. and Buehrle, J.},
   title = {Equilibrium drop surface profiles in electric fields},
   journal = {Journal of Physics: Condensed Matter},
   volume = {19},
   number = {37},
   pages = {375112},
   year = {2007}
}

@article{RN40,
   author = {Tsakonas, C. and Corson, L. T. and Sage, I. C. and Brown, C. V.},
   title = {Electric field induced deformation of hemispherical sessile droplets of ionic liquid},
   journal = {Journal of Electrostatics},
   volume = {72},
   number = {6},
   pages = {437-440},
   DOI = {10.1016/j.elstat.2014.08.004},
   year = {2014}
}

@article{corson2014,
   author = {Corson, L. T. and Tsakonas, C. and Duffy, B. R. and Mottram, N. J. and Sage, I. C. and Brown, C. V. and Wilson, S. K.},
   title = {Deformation of a nearly hemispherical conducting drop due to an electric field: Theory and experiment},
   journal = {Physics of Fluids},
   volume = {26},
   number = {12},
   pages = {122106},
   DOI = {10.1063/1.4903223},
   year = {2014}
}

@article{vancauwenberghe2013,
   author = {Vancauwenberghe, V. and Di Marco, P. and Brutin, D.},
   title = {Wetting and evaporation of a sessile drop under an external electric field: A review},
   journal = {Colloids and Surfaces A: Physicochemical and Engineering Aspects},
   volume = {432},
   pages = {50-56},
   year = {2013}
}

@article{RN38,
   author = {Sinn, Nico and Schuer, Maximilian T. and Hardt, Steffen},
   title = {No-contact electrostatic manipulation of droplets on liquid-infused surfaces: Experiments and numerical simulations},
   journal = {Applied Physics Letters},
   volume = {114},
   number = {21},
   pages = {213704},
   DOI = {10.1063/1.5091836},
   year = {2019}
}

@article{RN455,
   author = {Hartmann, J. and Schür, M. T. and Hardt, S.},
   title = {Manipulation and control of droplets on surfaces in a homogeneous electric field},
   journal = {Nature Communications},
   volume = {13},
   number = {1},
   pages = {27879},
   DOI = {10.1038/s41467-021-27879-0},
   year = {2022}
}

@article{RN5,
   author = {Pillai, Dipin S. and Sahu, Kirti Chandra},
   title = {Universal scaling law for electrified sessile droplets on a lyophilic surface},
   journal = {Physical Review E},
   volume = {109},
   number = {1},
   pages = {L013101},
   DOI = {10.1103/PhysRevE.109.L013101},
   year = {2024}
}

@article{RN10,
   author = {Langemann, Dirk},
   title = {A droplet in a stationary electric field},
   journal = {Mathematics and Computers in Simulation},
   volume = {63},
   number = {6},
   pages = {529-539},
   DOI = {10.1016/S0378-4754(03)00062-4},
   year = {2003}
}

@article{RN33,
   author = {Beroz, J. and Hart, A. J. and Bush, J. W. M.},
   title = {Stability Limit of Electrified Droplets},
   journal = {Physical Review Letters},
   volume = {122},
   number = {24},
   pages = {244501},
   DOI = {10.1103/PhysRevLett.122.244501},
   year = {2019}
}

@article{RN18,
   author = {Vo, Q. and Tran, T.},
   title = {Critical Conditions for Jumping Droplets},
   journal = {Physical Review Letters},
   volume = {123},
   number = {2},
   pages = {024502},
   DOI = {10.1103/PhysRevLett.123.024502},
   year = {2019}
}

@article{RN26,
   author = {Cavalli, A. and Preston, D. J. and Tio, E. and Martin, D. W. and Miljkovic, N. and Wang, E. N. and Blanchette, F. and Bush, J. W. M.},
   title = {Electrically induced drop detachment and ejection},
   journal = {Physics of Fluids},
   volume = {28},
   number = {2},
   pages = {022101},
   DOI = {10.1063/1.4940213},
   year = {2016}
}

@article{RN34,
   author = {Collins, R. T. and Jones, J. J. and Harris, M. T. and Basaran, O. A.},
   title = {Electrohydrodynamic tip streaming and emission of charged drops from liquid cones},
   journal = {Nature Physics},
   volume = {4},
   number = {2},
   pages = {149-154},
   DOI = {10.1038/nphys807},
   year = {2008}
}

@article{RN24,
   author = {Choi, K. and Ng, A. H. C. and Fobel, R. and Wheeler, A. R.},
   title = {Digital Microfluidics},
   journal = {Annual Review of Analytical Chemistry},
   volume = {5},
   pages = {413-440},
   DOI = {10.1146/annurev-anchem-062011-143028},
   year = {2012}
}

@article{RN11,
   author = {Zhu, Pingan and Wang, Liqiu},
   title = {Passive and active droplet generation with microfluidics: a review},
   journal = {Lab on a Chip},
   volume = {17},
   pages = {34-75},
   DOI = {10.1039/C6LC01018K},
   year = {2016}
}

@article{RN32,
   author = {Park, J. U. and Hardy, M. and Kang, S. J. and Barton, K. and Adair, K. and Mukhopadhyay, D. K. and Lee, C. Y. and Strano, M. S. and Alleyne, A. G. and Georgiadis, J. G. and Ferreira, P. M. and Rogers, J. A.},
   title = {High-resolution electrohydrodynamic jet printing},
   journal = {Nature Materials},
   volume = {6},
   number = {10},
   pages = {782-789},
   DOI = {10.1038/nmat1974},
   year = {2007}
}

@article{RN31,
   author = {Fenn, J. B. and Mann, M. and Meng, C. K. and Wong, S. F. and Whitehouse, C. M.},
   title = {Electrospray ionization for mass spectrometry of large biomolecules},
   journal = {Science},
   volume = {246},
   number = {4926},
   pages = {64-71},
   DOI = {10.1126/science.2675315},
   year = {1989}
}

@article{RN1,
   author = {Xiao, Ke and Wu, Chen-Xu},
   title = {Droplet dynamics driven by electrowetting},
   journal = {Physical Review E},
   volume = {105},
   number = {6},
   pages = {064609},
   DOI = {10.1103/PhysRevE.105.064609},
   year = {2022}
}

@article{xiao2021,
   author = {Xiao, Ke and Chen, Xi and Wu, Chen-Xu},
   title = {Electric field-triggered Cassie-Baxter-Wenzel wetting transition on textured surface},
   journal = {Physical Review Research},
   volume = {3},
   number = {3},
   pages = {033277},
   DOI = {10.1103/PhysRevResearch.3.033277},
   year = {2021}
}

@article{RN2,
   author = {Li, Jing and Wen, Kaiqiang and Xiao, Ke and Chen, Xiaoming and Wu, Chen-Xu},
   title = {Scaling law for the critical voltage of a droplet on a surface in the presence of an external field},
   journal = {Journal of Applied Physics},
   volume = {137},
   number = {5},
   pages = {054701},
   DOI = {10.1063/5.0245338},
   year = {2025}
}

@book{landau8,
   author = {Landau, L. D. and Lifshitz, E. M.},
   title = {Electrodynamics of Continuous Media},
   publisher = {Pergamon Press},
   address = {Oxford},
   edition = {2},
   year = {1984}
}

\end{document}